\documentstyle[aps,prb,epsf]{revtex}
\draft
\begin{document}

\title{$\kappa-(BEDT-TTF)_2X$ organic crystals:  superconducting versus 
antiferromagnetic instabilities in an anisotropic triangular lattice 
Hubbard model}

\author{Shan-Wen Tsai\footnote{new address: Department of Physics, University 
of Florida, Gainesville, FL 32611-8440 (tsai@phys.ufl.edu)} and J. B. Marston}

\address{Department of Physics, Brown University, Providence, RI 02912-1843}

\maketitle

\begin{abstract}
A Hubbard model at half-filling on an anisotropic triangular lattice has been 
proposed as the minimal model to describe conducting layers of 
$\kappa-(BEDT-TTF)_2X$ organic materials.  The model interpolates between 
the square lattice and decoupled chains.  The $\kappa-(BEDT-TTF)_2X$ 
materials present many 
similarities with cuprates, such as the presence of unconventional metallic 
properties and the close proximity of superconducting and antiferromagnetic 
phases.  As in the cuprates, spin fluctuations are expected to play a 
crucial role in the onset of superconductivity.  We perform a weak-coupling 
renormalization-group analysis to show that a superconducting instability 
occurs.  Frustration 
in the antiferromagnetic couplings, which arises from the underlying 
geometrical arrangement of the lattice, breaks the perfect nesting of 
the square lattice at half-filling.  The spin-wave instability is suppressed 
and a superconducting instability predominates.  For the 
isotropic triangular lattice, there are again signs of long-range magnetic 
order, in agreement with studies at strong-coupling.
\end{abstract}

\section{Introduction}
\label{sec:Intro}

The family of $\kappa-(BEDT-TTF)_2X$ layered organic crystals exhibit many 
fascinating electronic properties\cite{Williams,Ishiguro,McKenzie1}.
There are similarities with the high-$T_c$ cuprates\cite{McKenzie2}.  
Competition between 
antiferromagnetic and superconducting instabilities, seen in the cuprates, 
also appears in the $\kappa-(BEDT-TTF)_2X$ compounds.

\begin{figure}
\centerline{\epsfxsize=1.5in \epsfbox{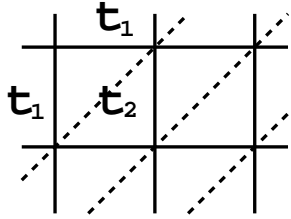}}
\caption{Anisotropic triangular lattice with two hopping amplitudes $t_1$ 
and $t_2$.} 
\label{fig:tri}
\end{figure}

The Hubbard model at half-filling on an anisotropic triangular lattice was 
proposed by McKenzie\cite{McKenzie1} as a model of the conducting layers of 
$\kappa-(BEDT-TTF)_2X$.  It is a simplification of a model first 
introduced by Kino and Fukuyama\cite{Kino1}.  Two hopping matrix elements 
are considered, $t_1$ between sites on a square lattice and 
$t_2$ between next-nearest-neighbors along one of the two diagonal directions 
(see Fig. \ref{fig:tri}).
The case $t_1 = t_2$ thus corresponds to the isotropic triangular lattice.  The 
model interpolates between the square lattice ($t_2 = 0$), 
which has been the subject of many studies in the context of high-$T_c$ 
cuprate superconductors and which has a spin-wave instability at half-filling, 
and completely decoupled chains ($t_1 = 0$) for which we have an exact 
Bethe ansatz solution by Lieb and Wu\cite{Lieb}.  The Hamiltonian is given by:
\begin{eqnarray}
H = -t_1 \sum_{<{\bf ij}>} (c_{\bf i}^{\dagger \sigma} c_{{\bf j} \sigma} + 
h.c.) - t_2 \sum_{<<{\bf ij} >>} (c_{\bf i}^{\dagger \sigma} c_{{\bf j} 
\sigma} + h.c.) + ~U_0 \sum_{\bf i} (n_{{\bf i} 
\uparrow}-\frac{1}{2})(n_{{\bf 
i} \downarrow}-\frac{1}{2}) + \mu \sum_{\bf i} n_{\bf i},
\label{eq:Hamiltonian}
\end{eqnarray}
where $<{\bf ij}>$ denotes nearest-neighbor pairs of sites on the square 
lattice and $<<{\bf ij}>>$ are next-nearest-neighbor pairs along one of the
two diagonal directions as shown in Fig. \ref{fig:tri}.

\begin{figure}
\centerline{\epsfxsize=6in \epsfbox{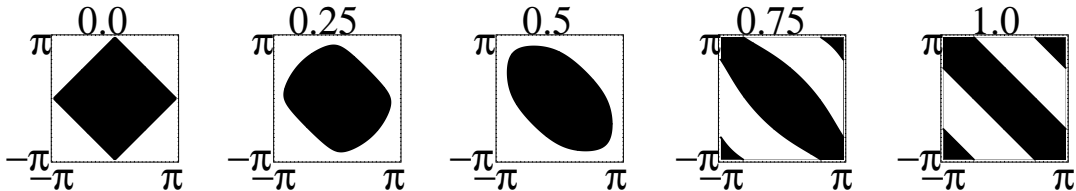}}
\caption{Fermi surface of non-interacting electrons for different values of 
the hopping amplitudes.  The number on the top of each graph is the 
value of the quantity $t_2/(t_1 + t_2)$, ranging from $0$ (square lattice) 
to $1$ (decoupled chains).  The chemical potential $\mu$ is varied to
ensure that the system is at half-filling.}
\label{fig:fs}
\end{figure}

Values for the hopping amplitudes, obtained from several quantum 
chemistry calculations\cite{hoppings1,hoppings2,hoppings3} and for 
different anions $X$, are in the range $t_1 > t_2$; that is, somewhere 
between the square and the isotropic triangular lattices.  
The non-interacting part of the Hamiltonian can be easily diagonalized and 
the corresponding Fermi surfaces at half-filling are depicted in 
Fig. \ref{fig:fs}.  

In the next section we present results from a weak-coupling 
renormalization-group analysis on this model.  We use the approach of 
Zanchi and Schulz\cite{Zanchi}.  The Fermi surface is divided 
into 16 patches.  One-loop RG flow equations for all the possible 
scattering processes involving the patches on the Fermi surface are numerically 
integrated towards the low-energy limit.
All subleading non-logarithmic contributions, arising from six-point 
functions generated during the mode elimination, are included. We conclude 
with a discussion of our results and make a comparison with 
results obtained in the strong-coupling limit.

\section{Results}
\label{sec:results}

\begin{figure}
\centerline{\epsfxsize=4in \epsfbox{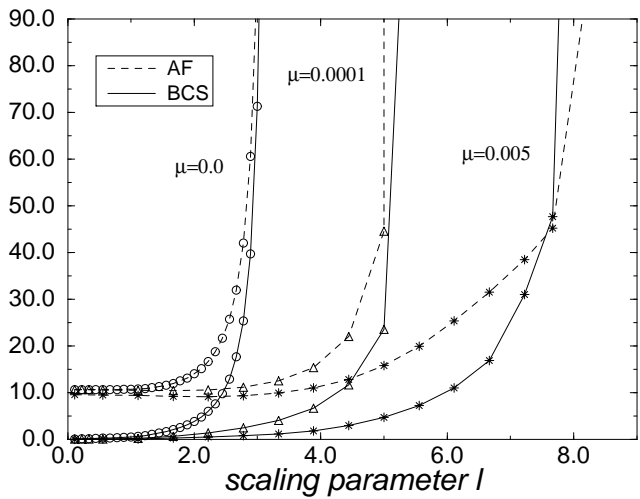}}
\caption{Results for the square lattice ($t_2 = 0$).  At half-filling 
($\mu = 0$), spin-wave instability occurs, 
but as the system is doped away from half-filling, the perfect nesting is 
lost and there is a crossover to a BCS instability.}
\label{fig:afxsc}
\end{figure}

The Hubbard model on the pure square lattice ($t_2 = 0$) and with repulsive 
on-site interaction $U_0 > 0$ has been extensively 
studied.  Weak-coupling renormalization-group analyses of the couplings,
resolved into discrete patches along the Fermi surface, have been 
performed\cite{Zanchi,Halboth1,Honerkamp}.  At half-filling, since the 
Fermi surface is fully nested (Fig. \ref{fig:fs}), an
antiferromagnetic (AF) spin-wave instability develops.  As the 
system is doped slightly away from half-filling, there is a crossover to a
superconducting BCS regime and the dominant
Cooper pairing symmetry is $d_{x^2-y^2}$.  As shown in Fig.
\ref{fig:afxsc}, we have reproduced these results\cite{Zanchi,Halboth1}.  

\begin{figure}
\centerline{\epsfxsize=4in \epsfbox{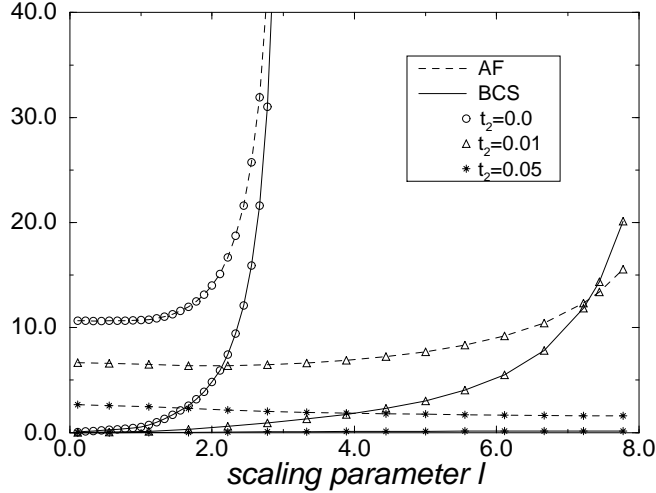}}
\caption{At half-filling, spin-wave instability occurs for $t_2 = 0$, but 
as $t_2$ increases, the BCS instability wins over.  Finally, as $t_2$ is 
increased further, the nesting is destroyed and both divergences are 
suppressed.}
\label{fig:afxsc_t2}
\end{figure}

Introducing non-zero $t_2$ offers a {\it different} way of favoring a BCS 
instability over AF.  Perfect nesting is eliminated once the hopping $t_2$ 
between next-nearest-neighbors sites along one of the two diagonal directions
is turned on (Fig. \ref{fig:tri}).  In contrast to 
the square lattice where both AF spin-wave and BCS couplings 
increase during the renormalization-group transformations with $V^{AF}$ 
diverging faster than $V^{BCS}$, for sufficiently large $t_2$ there is a 
crossover to a regime where the BCS processes eventually dominate, 
signaling a superconducting instability. 
Furthermore, because the Fermi surface is imperfectly nested, the growth 
in both types of couplings weakens.  Further increasing $t_2$ 
eventually destroys the nesting of the Fermi surface altogether and both 
types of divergences are suppressed.  Three cases illustrating these 
crossovers are shown in Fig. \ref{fig:afxsc_t2}.  

\begin{figure}
\centerline{\epsfxsize=4in \epsfbox{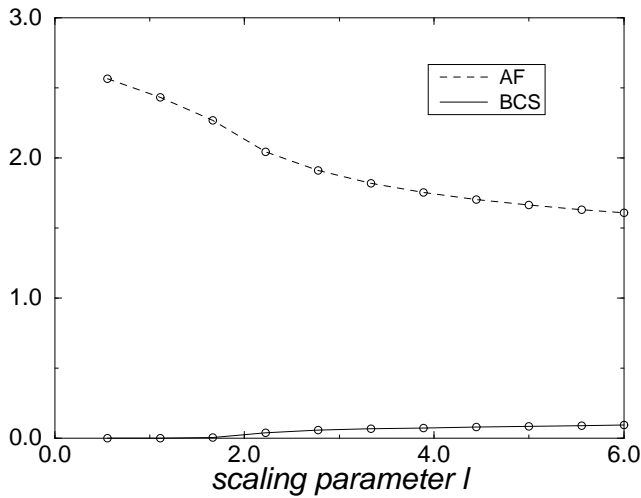}}
\caption{Flow of dominant AF and BCS channels for the isotropic triangular 
lattice.}
\label{fig:afxsc_tri}
\end{figure}

Going further to the isotropic point, our weak-coupling RG analysis no
longer shows BCS instabilities.  In Fig. \ref{fig:afxsc_tri} 
the dominant AF and BCS channels are compared.  Neither channel shows strong
divergences, but the AF channel is significantly larger than the BCS 
channel.  Slave-boson calculations\cite{Gazza} find a Mott-Hubbard 
metal-insulator transition to occur at the relatively large value of 
$U_c = 7.23 t$, but also find a metallic phase with incommensurate spiral 
order\cite{Capone} at $U < U_c$.  Our weak-coupling results seem to indicate 
the onset of re-entrant long-range antiferromagnetic order.
AF ordering tendencies again disappear as $t_2$ is increased further, beyond
$t_1$.  This is as expected, since decoupled Hubbard chains do not
exhibit AF order.  

\section{Conclusion}
\label{sec:Conclusion}

Hubbard models on the square lattice have been extensively studied in the 
context of high-$T_c$ superconductivity.  We have reproduced the well-known
result that there is an AF instability at half-filling.  Also doping the system 
away from half filling induces a crossover to a BCS regime with $d_{x^2-y^2}$ 
pairing symmetry, as expected.  More importantly, we have shown another way 
of triggering a BCS instability.  Keeping the system 
at half-filling, but introducing the diagonal hopping $t_2$ (as shown in Fig. 
\ref{fig:tri}), eliminates perfect nesting.  Corresponding magnetic 
frustration kills the spin density wave, and Cooper pairing can dominate.
This result suggests that superconductivity may occur in a model of 
strongly-correlated electrons that interpolates between the square and 
isotropic triangular lattices.

At the isotropic point, there are signs of re-entrant antiferromagnetic 
long range order.  In the large-$U_0$ limit, at half-filling, 
the Hubbard model can of course
be mapped to the spin-1/2 Heisenberg antiferromagnet which
is known to have long-range AF order on the triangular lattice. 
Furthermore, away from the isotropic point, our weak-coupling RG analysis 
continues to agree qualitatively with results obtained for the pure Heisenberg
system with two exchange couplings $J_1$ and $J_2$.  The phase diagram of the
Heisenberg model has been studied via a straight $1/S$ expansion\cite{Merino}, 
series expansion methods\cite{Weihong}, and 
a large-N treatment\cite{Chung}.  All three methods find two regions of 
long-range order: near the limit of a square lattice ($J_2 = 0$) and near the 
isotropic point ($J_2 = J_1$).  It is remarkable that our weak-coupling
analysis agrees with these results.

{\bf Acknowledgments}
We thank Chung-Hou Chung, Anthony Houghton, Ross McKenzie, Jaime Merino
and Matthias Vojta for helpful discussions.

\end{document}